\documentclass[aps,pra,twocolumn,showpacs]{revtex4}
\usepackage{epsfig,latexsym,amssymb}

\begin{document}

\title{Metastable level lifetimes from electron-shelving measurements\\
 with ion clouds and single ions }
\author{Martina Knoop}
\author{Caroline Champenois}
\author{Ga\"etan Hagel}
\author{Marie Houssin}
\author{Caroline Lisowski}
\author{Michel Vedel}
\author{Fernande Vedel}
\affiliation{ Physique des Interactions
Ioniques et Mol\'eculaires, CNRS - Universit\'e de Provence (UMR 6633)\\
Centre de Saint J\'er\^ome, Case C21, 13397 Marseille Cedex 20,
France.\\}
\email{mknoop@up.univ-mrs.fr}

\date{\today}


\begin{abstract}The lifetime of the $3d^2D_{5/2}$-level in
singly-ionized calcium has been measured by the electron-shelving
technique on different samples of rf trapped ions. The metastable
state has been directly populated by exciting the
dipole-forbidden $4S_{1/2} - 3D_{5/2}$ transition. In ion clouds,
the natural lifetime of this metastable level has been measured
to be (1095$\pm$27) ms. For the single-ion case, we determined a
lifetime of (1152$\pm$20) ms. The $1\sigma$-error bars at the
2\%-level have different origins for the two kinds of experiments:
data fitting methods for lifetime measurements in an ion cloud
and control of experimental parameters for a single ion.
De-shelving effects are extensively discussed. The influence  of
differing approaches for the processing of the single-ion quantum
jump data on the lifetime values is shown. Comparison with recent
measurements shows excellent agreement when evaluated from a
given method.
\end{abstract}

\maketitle

\section{Introduction}
\label{s-introduction} High-precision atomic lifetime
measurements are a challenge for experimental and theoretical
atomic physics. Increasing volume and precision of astrophysical
observations have generated a growing need for precise atomic
lifetime data \cite{traebert00}. Among all the interesting atomic
systems, atoms with a closed inner shell and a single additional
valence electron offer the possibility to accurately compare the
results of atomic data modeling and experimental values.

There have been various theoretical approaches to the low-lying
metastable doublet levels of $^{40}$Ca$^+$ which have generated
somewhat dispersed metastable lifetime predictions
\cite{osterbrock51,warner68,ali88,zeippen90,guet91,vaeck92,brage93,liaw95,biemont96}.
 In the past decade the lifetimes of the metastable D-levels of the
  singly-ionized calcium have also been measured several times
\cite{urabe92,arbes93,urabe93,arbes94,knoop95,gudjons96,ritter97,block99,lidberg99,barton00,staanum03}.
Most of these measurements have been carried out 
in rf traps, showing a clear evolution towards longer lifetimes
throughout the years, as error sources have been identified. In
particular, before 1999, the coupling of the metastable D-levels
to the P-state due to off-resonant excitation by the repumper
laser \cite{block99}, had not been taken into account at all. As
will be shown in this paper, ion-ion collisions in a relative hot
and dense ion cloud ($k_BT\simeq1eV$, $n\simeq10^8cm^{-3}$) may
also contribute a few percent to the lifetime reduction. These
two effects  explain the gap between the earlier ion-cloud
measurements and recent single-ion measurements.
 The experiments which have been carried out on single laser-cooled ions show a
typical error bar of the order of 2\%, with the exception of one
experiment \cite{barton00}, where the announced uncertainty is
inferior to 0.6\% and which gives the highest value ever measured
for the $3D_{5/2}$-lifetime.
In this article, we  present our new
measurements of the lifetime of the $3D_{5/2}$-state which have
been carried out in a different way. In fact, we achieve
population of the metastable level by direct pumping of the
electric quadrupole transition which connects the ground state to
the $3D_{5/2}$-state. We have also been able to establish a
connection between the different published values, the
discrepancy being mainly due to differing data analysis.

 Single trapped ions are now extensively used for metrology and
quantum information \cite{Gill2002}. Storage times can exceed
days, and the control of the trapping environment along with the
degree of laser stabilization have been improved to access
linewidths of optical transitions down to the Hz-level
\cite{rafac00}. Our  experimental project aims to propose a
frequency standard in the optical domain, based on the
interrogation of the $4S_{1/2}-3D_{5/2}$ electric quadrupole
transition of a single laser-cooled Ca$^+$-ion. The measurement
of the metastable $3D_{5/2}$-level lifetime is an important step
in this direction, allowing to identify and to control effects
which may reduce this lifetime, and which could thus contribute
to the broadening of the clock transition.

In this paper, we  briefly present our experimental device, and
 introduce the technique we have used to measure metastable
lifetimes in ion clouds and single ions. We then
  present the different data processing methods. The fourth part of this article is
devoted to the discussion of the uncertainties of our
measurements. We finally present and discuss our results in the
context of previous experiments.


\section{Experimental Setup}

\label{s-expsetup} 
\begin{figure}[here]
\begin{center}
   \epsfig{file=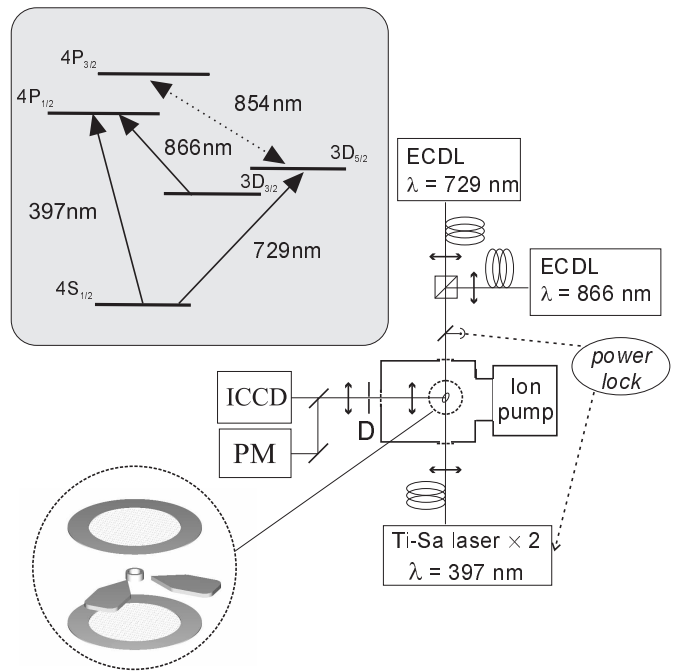, width=90mm}
    \end{center}
  \caption{Experimental setup of the miniature ion trap.
   The insets show the Ca$^+$ lowest lying levels and the geometry
   of the miniature trap surrounded by the four compensation electrodes.}
  \label{f-setup}
\end{figure}

We use a miniature ion trap to confine single
ions and ion clouds up to approximately 500 particles. The trap is
a modified Paul trap, a so-called Paul-Straubel trap
\cite{schrama93} consisting of a cylindrical molybdenum ring with
an inner diameter of $2r_0=1.4$ mm and a total ring height of
$2z_1=0.85$ mm  (Figure \ref{f-setup}).  Two circular mesh
electrodes at 5.5 mm on each side from the trap center, allow the
definition of well controlled electrical boundary conditions. Two
copper tip electrodes in the plane of the trapping ring are used
as positioning electrodes to correct for imperfections and
asymmetries in the trapping potential and thus to reduce the
micromotion of the confined ions. The applied trapping frequency
$\Omega/2\pi=11.6$ MHz with an rf voltage amplitude of  300 V,
gives rise to a total pseudo-potential well depth of $1.9$ eV and
motional frequencies around $\omega_i/2\pi=1.5$ MHz. The trap has
been fully characterized and is described in detail in
\cite{champenois01}.
The  trapping setup is mounted into an ultra-high vacuum vessel
and baked out  at $150^{\circ}$C during a couple of days. Vacuum
conditions can be controlled by the ion pump current, a
Bayard-Alpert gauge and a mass spectrometer operating up to 64
a.m.u.. Ions in the trap are created from a calcium oven heated
by a direct current of 3 A. The effusive atom
beam is crossed with a low-energy electron beam 
in the center of the trap producing singly-ionized calcium ions.

Laser-cooling is carried out on the $4S_{1/2}-4P_{1/2}$ electric
dipole transition at 397 nm using an intracavity
frequency-doubled TiSa-laser [\textsl{Coherent 899}]. The output
intensity of this laser is stabilized  by a single-pass
acousto-optical modulator (AOM), the error signal is fed back
from a beam pick-up photodiode after crossing the trap. The
optical power used for laser-cooling is about 50 $\mu$W in the
case of an ion cloud, and an order of magnitude lower for a single
ion, focussed in a 20$\mu$m-diameter waist (measured at 1/e$^2$
power level). The linewidth of this laser has been found to be
below 10 MHz in the course of our experiments.
Due to a branching ratio to the $3D_{3/2}$-level larger than 5\%,
repumping from this level is required, and is assured by a
single-mode diode laser at 866 nm mounted in a
Littrow-configuration external cavity (ECDL). This diode laser is
stabilized to a low finesse ($\mathcal{F}$=200) reference cavity
reducing its linewidth to about 1 MHz and improving its frequency
stability. Stable operation during a whole day is achieved by
locking the length of the reference cavity to a hyperfine
transition of neutral caesium making use of an additional ECDL at
852 nm and a standard saturated-absorption setup. Typical power
for the 866 nm-diode is about 200 $\mu$W focussed into a 70
$\mu$m-diameter waist.
A broad-area laser diode is used to probe the electric quadrupole
transition $4S_{1/2}-3D_{5/2}$ at 729 nm . The nominal output
power of this  laser  diode is 100 mW and its free-running
linewidth is as large as 2 nm. This diode has been set up in an
external cavity in Littrow configuration and pre-stabilized by
electronic feedback onto a low-finesse reference cavity
($\mathcal{F}$=200) . We thus obtain a laser linewidth in the
MHz-range and a laser intensity of approximately 350 $\mu$W per
60 $\mu$m-diameter waist size in the trap \cite{houssin03}.

Two mechanical shutters in the 729 nm-beam line allow to cut the
light from this laser completely. Their closing time is in the
ms-range, inferior to the smallest measurement interval used
throughout the experiment.
All the laser beams used in this experiment are brought to the
ion trap by single-mode optical fibers. This ensures the spatial
filtering of the laser beams and gives rise to well controlled
waist sizes at the position of the ion. However,  the main
advantage of these fibers is to increase the pointing stability
of the laser beams, which improves the day-to-day reproducibility
of the measurements essential for trapping  single ions.

Detection of the fluorescence of the ions at 397 nm is made outside the
vacuum vessel (Figure \ref{f-setup}). The fluorescence signal is
spatially filtered by a  small diaphragm (diameter below 500
$\mu$m) and then projected onto an intensified CCD camera (ICCD)
and a photomultiplier (PM) in photon counting mode. A variable
beamsplitter between these two devices allows to choose the
fraction of signal sent to the photomultiplier between 10 and 100
\%. The maximum fluorescence count is about 10000 counts/sec for
a single ion.
Data is collected and stored by a personal computer, that also
controls the ion creation process and actuates the laser shutters.
Excellent temporal synchronization is assured by
buffer-controlled read-out on the data acquisition board. The
typical duration of a measurement bin is 50 ms for the ion cloud
measurements and 30 ms for the single-ion measurements.

\section{electron shelving}
All our lifetime measurements have been performed in the same
trap, in a very similar way  for a small ion cloud or a single
ion. In both cases, ions have been laser-cooled on the strong
$4S_{1/2}-4P_{1/2}$ electric dipole transition and detected by
 the scattered 397 nm-photons. We have used
the technique of electron-shelving proposed by H.G. Dehmelt
\cite{dehmelt75} for the whole set of measurements. This method
allows to observe transitions on the forbidden electric
quadrupole line via the switching of the strong laser-cooling
transition at 397 nm. In our experiment the $3D_{5/2}$-state has
been populated by direct optical pumping of the dipole-forbidden
$4S_{1/2}-3D_{5/2}$ transition, technique which has not yet been
applied in the lifetime measurements with single trapped
Ca$^+$-ions. The direct excitation of this electric quadrupole
transition allows an unambiguous definition of initial
conditions.

\subsection{ measurements in an ion cloud}
\label{s-cloud} A typical ion cloud contains 50 to 100
Ca$^+$-ions which are  laser-cooled to a temperature of about 50
K, temperature which is defined as a measure of the kinetic
energy of the trapped particles. This temperature value stands
for an equilibrium between the heating of the ions by the
trapping field and the  laser-cooling.

\begin{figure}
\begin{flushleft}
    \epsfig{file=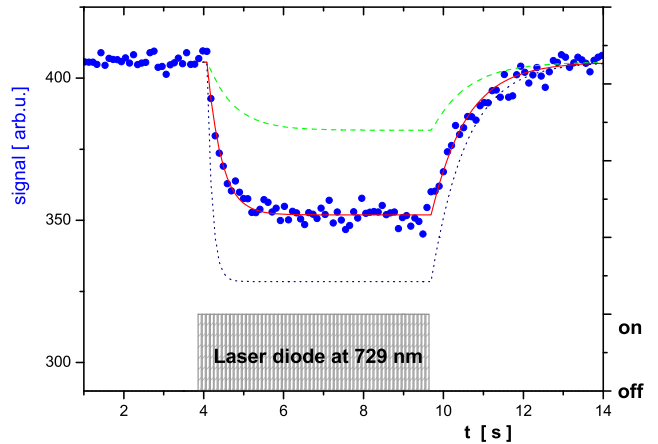, width=95mm}
      \end{flushleft}
  \caption{electron-shelving data in an ion cloud. The applied 729-nm laser power has been recorded simultaneously and
   is shown in the lower trace. The lines show simulation data by the density matrix  formalism with various
   power levels  for the probing laser at 729 nm (0.25 $I_0$ (dashed), $I_0$ (solid), 4 $I_0$ (dotted)) }
  \label{f-shelving}
\end{figure}

To measure the $3D_{5/2}$-level lifetime in an ion cloud  we use
approximately 350 $\mu$W of   729 nm-power. Once the resonant laser
beam is admitted onto the ions, a fraction of the ions is pumped
into the $3D_{5/2}$-state and the observed fluorescence signal at
397 nm  decreases. The experimental signal is shown in figure
\ref{f-shelving}. When the  population of the $3D_{5/2}$-state
has reached an equilibrium state which is represented by a
constant fluorescence level, we shut off the 729 nm diode laser.
The ion population then decays to the ground state with a time
constant governed by the $3D_{5/2}$-state natural lifetime. This
decay can be observed on the 397 nm fluorescence signal as the
decaying ions return into the laser-cooling cycle and scatter
blue photons. The revival function of the 397 nm fluorescence is
fitted with a least-squares method (LSF) by the function $ F(t) =
S_0 + S \cdot (1- exp(-t/\tau_m))$ where $\tau_m$ is the measured
lifetime for a given set of experimental parameters, $S$ the net
ion signal and $S_0$  the low-level signal being composed by the
fluorescence signal of the ions remaining in the laser-cooling
cycle and the background signal due to stray light. Actually, the
measured lifetime $\tau_m$ is the natural lifetime of the
$3D_{5/2}$-level affected by different de-shelving effects as
will be discussed in section \ref{s-eval-err}. The result for the
measured lifetime in an ion cloud is $\tau_{nat}=1095$ ms with an
1$\sigma$-statistical uncertainty of 7.5 ms.

We have made simulations in  an ion cloud of the population of
all the atomic levels involved ($S_{1/2}$, $P_{1/2}$, $D_{3/2}$,
$D_{5/2}$)  with the use of the density matrix formalism taking
into account the ion motion in the trap. The oscillatory movement
of the ion cloud is described by a distribution of the velocity
amplitudes at a single frequency \cite{schubert89}. In this
coupled four-level system, the fraction of ions which are pumped
into the metastable $D_{5/2}$-state depends on the relative
detuning of the three lasers used for cooling, repumping and
probing of the clock transition, a phenomenon comparable to the
occurrence of dark resonances. The simulation  shows, that it
is possible to pump almost the entire ion population into the
metastable state if the ions are nearly at rest \cite{champenois04}. The second
parameter which is critical for the reproduction of the shelving
curves is of course the laser power. The lines in figure
\ref{f-shelving} visualize the simulated fluorescence curves for
a  cloud of 50 ions with a temperature of 100 K, assuming that
the fluorescence is directly proportional to the
$P_{1/2}$-population. In the plotted simulations the relative
laser detuning has been fixed while the  power levels of the
probing laser at 729 nm are varied.   The right choice of
relative laser detunings and absolute laser power levels enables
us to
 reproduce accurately both the speed of the decrease of the fluorescence as
ions leave the laser cooling cycle (t=4-9 sec) and the fraction
of the ions being pumped into the $D_{5/2}$-level.  As a matter of
course, the revival of the fluorescence depends only on the
lifetime of the 3$D_{5/2}$-state for the given set of
experimental parameters.

\subsection{ measurements with a single ion}
\label{s-single ion}

Experiments with single ions have been carried out during twelve
6h-runs with single ions laser-cooled to temperatures below 1K.
As the temperature of the ion is estimated from the asymmetric
Doppler profile which is largely depending on the optical power
used for laser-cooling, this value is an upper limit to the
estimate of the kinetic energy.

\begin{figure}
\begin{center}
   \epsfig{file=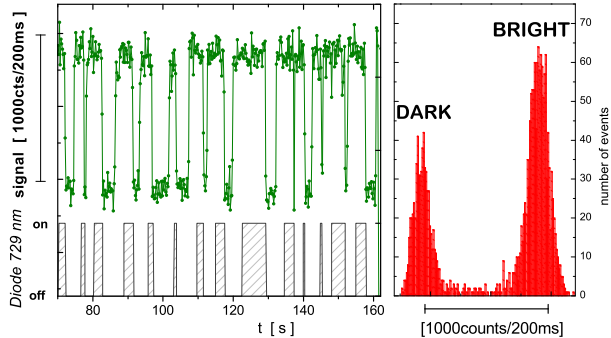, width=9cm}
  \end{center}
  \caption{\textit{left}: Quantum jumps of a single Ca$^+$ ion in the temporal evolution of the fluorescence signal, the lower trace
  shows the switching of the exciting diode laser at 729 nm;
   \textit{right}: Histogram of the fluorescence signal
   visualizing the distinction between the upper and the lower
   trace.}
  \label{f-qj}
\end{figure}

When the electron-shelving technique is applied to a single ion,
the observed fluorescence signal becomes binary. Actually, while
the ion is in the laser-cooling cycle, a large number of blue
photons is scattered giving rise to a high photon count rate
("bright" level), see figure \ref{f-qj}. When the 729 nm
laser diode is admitted onto the ion, it will be pumped into the $3D_{5/2}$-level.
The fluorescence rate at 397 nm will then abruptly fall to the
low level, which corresponds to the background light ("dark"
level). As soon as we detect a "dark" level signal, the 729-nm
laser is shut off to avoid any coupling between the levels, as
discussed in detail in section \ref{s-eval-err}. The sudden changes
in fluorescence due to transitions between atomic levels are
called quantum jumps. Figure \ref{f-qj} shows the good
distinction between the bright and the dark level. This allows to
define an unambiguous threshold value for the determination of
the duration of the quantum jumps. Typically, we have set the
threshold value at the half of the net signal value, which is the
difference between the mean upper trace and the mean lower trace.
The timebase of the data acquisition has been chosen to give a
maximum separation of bright and dark level together with best
temporal resolution. Thus, variation of the threshold value
separating dark and bright level between one third and two third
of the net signal did not result in variations of the quantum
jump length distribution.

\begin{figure}
\begin{center}
   \epsfig{file=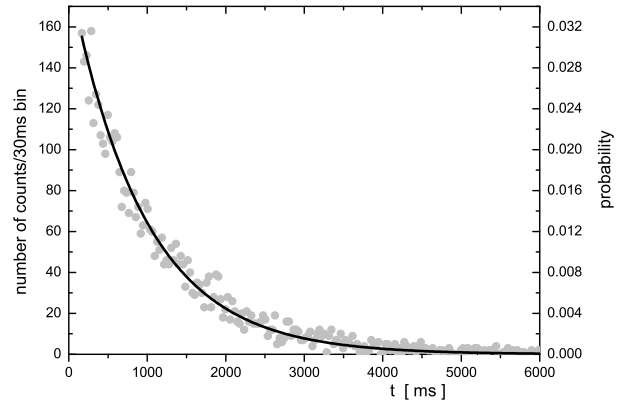, width=9cm}
  \end{center}
  \caption{Histogram of the duration of the measured dark intervals observed with a single ion. The left y-axis scale represents
  the absolute event counts, while the right scale indicates the probability of the event.}
  \label{f-qjhist}
\end{figure}

The duration of the dark intervals have been processed by two
different methods. First, they have been binned in a histogram
and fitted by a least-squares method (LSF) using an exponential
decay curve ($ F(t) = A \cdot exp(- t/\tau_m)$) with two free
parameters (A and $\tau_m$) at a 95\% confidence level (figure
\ref{f-qjhist}). Taking into account the de-shelving by the 866 nm-diode as
discussed in section \ref{s-eval-err}, the obtained lifetime value is
$\tau_{nat}$ = 1088 ms with a fitting uncertainty of 15 ms.
This fitting procedure assumes that the measured number of events
for each bin spreads around the fitted equation with a Gaussian
probability distribution. Actually, the probability of very long
events is in most cases very low (see figure \ref{f-qjhist}),
suggesting that deviations from the mean are not governed by a
Gaussian distribution. It is then appropriate to apply the most
general evaluation method, assuming a multinomial distribution,
and calculating the lifetime by a maximum likelihood estimate
(MLE). Recent measurements \cite{barton00,staanum03} have
evaluated the metastable lifetime by using the method of the MLE
obtaining  values larger than 1150 ms.
 We have used the following formula
\cite{staanum03} to calculate the lifetime by
\begin{equation}
\label{e-tau}
  \tau_m = \frac{\Delta t}{ln(1+\frac{\Delta t}{\overline{t}})}  ,
   \hspace{7mm} with  \hspace{7mm}\overline{t} = \frac{1}{N}\sum_{i=1}^N n_i t_i
\end{equation}
where $\Delta t$ is the bin size of the histogram, $\overline{t}$
is the mean value of the $n_i$ measured dark intervals of
duration $t_i$, and $N$ is the total number of measured events.
 The statistical
uncertainty for each data set can be exactly computed by

\begin{equation}
\label{e-vartau}
  \sigma =  \sqrt{\frac{\tau^2}{N}
   \left[1+\frac{1}{12}\left(\frac{\Delta t}{\tau}\right)^2+\mathcal{O}\left(\frac{\Delta t}{\tau}\right)\right]}
\end{equation}

Depending on the number of quantum jumps per run, the statistical
uncertainty has been found to be below 3\%. Extrapolation to zero
866nm-intensity gives a value for the natural lifetime of
$\tau_{nat}$ = 1152 ms, with a fitting uncertainty of 19 ms,
which is discussed in section \ref{s-dataan}.

\begin{figure}
\begin{center}
   \epsfig{file=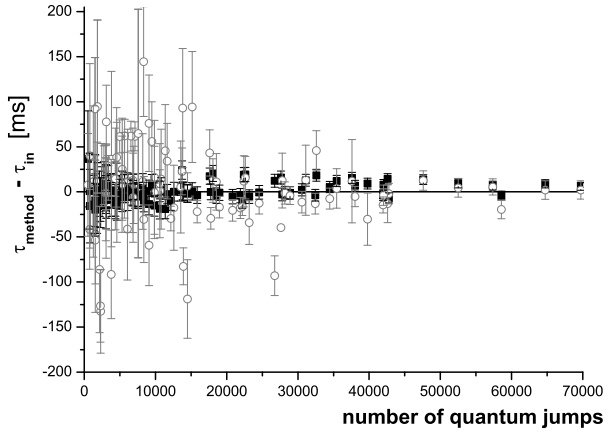, width=9cm}
  \end{center}
  \caption{Comparison of simulated lifetime data using MLE ($\blacksquare$) and LSF ($\bigcirc$) as
  a function of the number of quantum jumps. The graph shows the difference between the evaluated
   lifetimes and the lifetime injected into the simulation process ($\tau_{in}$). Both methods tend to
  produce identical results on a 2\%-level only starting around 40000 events.}
  \label{f-tausimul}
\end{figure}

To determine the appropriate method for the evaluation of the
quantum jump data, we have simulated quantum jump events using the
Monte-Carlo method, governed by a fixed input lifetime $\tau_{in}$
 and with the same
time base as in our experiment. These quantum jump lengths have been
allotted randomly in files of different size starting at about
500 quantum jumps up to about 100 000. Each file has been
processed by a least-squares fit to an exponential decay curve of
the histogram of durations and by the maximum-likelihood estimate
for the whole data set. The two data processing methods have been
carried out in a totally identical way to the treatment of the
experimental data. The obtained values are plotted in figure
\ref{f-tausimul} as a function of the number of quantum jump
events contained in the processed file. As can be seen, LSF data
are more widely scattered around the input lifetime of the
simulation. At about 10000 events the scattering of the MLE data
is roughly a factor of three smaller than for LSF, values for
both methods converge on a 2\%-level only for runs with more than
40000 valid quantum jumps. Actually, the MLE method is well known in
mathematics as a precise unbiased estimator, it reaches
asymptotically the Cram\'er-Rao bound which describes the best
attainable error \cite{Kendall61}. In general, as the average
time needed to record one quantum jump is of the order of two or
three seconds including the preparation time, a few-hour run of
data acquisition yields several thousands quantum jumps. In this
range the difference between both methods may easily reach 10\%
of the absolute values, the 5.5\%-discrepancy of our experimental
results largely fits into this window.  In summary, the number of
quantum jump events we 
acquire in the  course of a one-day run, imposes the use of MLE
to reach minimum statistical uncertainties.


\section{Evaluation of uncertainties}
\label{s-eval-err} Among the uncertainties on the measured
lifetimes  of the metastable $3D_{5/2}$-level of Ca$^+$ we
distinguish between the ones which are due to physical effects
which may affect the apparent lifetime of the state and those due
to  data analysis.

\subsection{De-shelving effects}

The lifetimes we measure in the course of our experiments are
function of the experimental parameters as various physical
effects tend to shorten the observed lifetimes. The major
de-shelving effects are collisions with the residual gas, heating
or loss of the ions, and coupling of the atomic levels by laser
light. The measured lifetime $\tau_m$ can therefore be expressed
as
\begin{equation}\label{eq-deshelv}
  \frac{1}{\tau_m} = \frac{1}{\tau_{nat}} + n_B
  (\Gamma_{quench}+\Gamma_{mix}) + \gamma_{heat} + \gamma_{loss} +
  \gamma_{coupling}.
\end{equation}
with $n_B$ the residual particle density in the trapping device.
 We will discuss the influence of these de-shelving effects in detail in the
following, their quantitative influence on the measured lifetimes
is summarized in table \ref{table_deshelv}.

\begin{table*}
  \centering
  \caption{Error budget for the measurements  of the $3D_{5/2}$ metastable lifetime. Data has been processed
 using least-squares fit (LSF) and multinomial   maximum-likelihood estimate
 (MLE).  The fitting uncertainty in the extrapolation
 of the quantum jump data includes the uncertainty on the power measurement.}
  \label{table_deshelv}
  \begin{tabular}{rccc}
 & \multicolumn{1}{c}{ion cloud measurement}& \multicolumn{2}{c}{quantum jump
 measurements}\\
   \hline
    \vspace{-1mm}
 evaluation method   &            LSF                &  LSF                         & MLE\\

   &         $\tau = 1095$ ms  &          $\tau = 1088$ ms &          $\tau = 1152$ ms \\
  \hline
 fitting uncertainty      &               7.5 ms                               &   15 ms  &   19 ms \\
 data analysis          &                    17.5 ms                      &   -      &   -     \\
 collisions             &      (2.1 $\pm$ 1.9)ms  and (16.8 $\pm$ 12) ms  &   3.5 ms &   3.5 ms\\
 heating                &                      1.1 ms                     &    -     &   -     \\
 ion loss               &                      8.4 ms                     &       -  &   -     \\

 \hline
  \hline
total error bar ($1\sigma$)      &               27 ms                    &    16 ms &    20 ms \\

  \end{tabular}
\end{table*}

We have looked for an eventual effect of  the trapping potential
and of the detuning of the 397 nm laser. Both parameters could
influence the temperature of the trapped ions and eventually give
rise to an increased amount of collisions. Nevertheless, we could
not evidence any variation of the measured lifetime in the limit
of the announced error bars.

\subsubsection{Collisional effects}
The apparent lifetime of the metastable state can be reduced by
inelastic collisions with (neutral) particles from the residual
gas background. The base pressure of the ultra-high vacuum vessel
is below $5\times 10^{-10}$ mbar as measured by the ion pump
current and the two gauges. From our previous measures of
quenching and j-mixing reaction rates in hot ion clouds
\cite{knoop95,knoop98} and from the measured composition of the
residual gas background by the mass spectrometer, we can deduce
that the only non-negligible component giving rise to inelastic
collisions is hydrogen. In fact, the partial hydrogen pressure
for ion cloud experiments is inferior to $2\times 10^{-9}$ mbar,
and below the resolution of the mass spectrometer ($1\times
10^{-9}$ mbar) for single ion experiments. The difference in the
pressure values for both types of experiments is due to the
duration and frequency of the ion creation process.

For the ion cloud experiments the residual hydrogen pressure
results in a value of
 $n_B \Gamma_{quench} \leq (1.8 \pm 0.7)\times 10^{-3}$ s$^{-1}$ due
 to quenching while the contribution of the fine-structure mixing
 collisions is
$n_B \Gamma_{mix} \leq (14 \pm 10)\times 10^{-3}$ s$^{-1}$.

The observation of quantum jumps in a single ion allows the
direct measurement of the collision rate in the absence of the
729-nm laser. Every once in a while, the ion undergoes a quantum
jump due to an inelastic collision with a particle from the
residual gas. We have recorded these quantum jumps which are the
result of collisional transfer between the fine-structure levels
and from the ground-state. At a base pressure of $2 \times
10^{-10}$ mbar we have measured an average of 1 quantum jump
every 5 minutes resulting in a collision rate of $n_B
\Gamma_{coll} \leq 3\times 10^{-3}$ s$^{-1}$.

\subsubsection{Ion heating}
Temperature changes in the ion cloud could play a role in the
determination of lifetimes due to changing collision rates and to
varying frequency overlap (and therefore varying excitation
efficiency) between the ions and the applied lasers. In the ion
cloud experiments we have adjusted the 729-nm laser power to
make sure that not more than one half of the ion cloud is pumped
into the metastable state. As the laser-cooling is permanently
applied to the rest of the ion cloud, sympathetic cooling of the
dark ions prevents these from heating \cite{larson86}. We can yet
estimate the influence of ion heating from the signal variations
at fixed laser frequencies, the upper limit  is
$\gamma_{heat} \leq 1\times 10^{-3}$ s$^{-1}$.

As for the single ions, heating becomes visible if the
laser-cooling parameters (mainly the frequency and the power of
the 397-nm laser) are largely detuned. After a dark interval the
ion will then slowly return to the "bright" fluorescence level
over several measurement bins. In our lifetime measurements, we
have made sure that the laser-cooling parameters are optimized and
that the ion signal returns to the upper fluorescence trace in a
time shorter than one measurement bin (30 ms), even for dark times
which are superior to 10 seconds. Moreover, to ensure efficient
cooling of the ion, we maintain it in the laser-cooling cycle for
at least 600 ms  before the 729 nm laser is again admitted to the
trap.

\subsubsection{Ion loss}

The revival function of the fluorescence in the ion cloud could
be biased by competing decay processes, in particular ion loss.
Ion loss is not visible on a single revival function but it will
be detectable on the over-all fluorescence signal after a one-hour
run. The upper limit for the influence of ion loss on the
lifetime measurements in an ion cloud is given by  $\gamma_{loss}
< 7\times 10^{-3}$ s$^{-1}$.

If the ion is lost, while we are observing quantum jumps, we have
to reload a new ion. To make sure, that the vacuum conditions are
identical throughout the whole set of experiments, we wait for
about half an hour before we restart  data acquisition.

\subsubsection{Level coupling by laser light}

All the laser beams used in this experiment may
couple the different atomic levels due to the Stark effect
induced by their electromagnetic field. However, atomic level
coupling by the 397 nm laser is negligible in our experiment, laser power at this
wavelength being around 50 $\mu$W for an ion cloud and below 10
$\mu$W for the experiments with a single ion.

\paragraph{729 nm}

 One important factor for the reduction
of the metastable lifetime is coupling by the 729-nm laser. In
fact, even at low laser powers, this radiation may induce transitions
to the ground state by stimulated emission on the
$4S_{1/2}-3D_{5/2}$ line. We have taken precautions to avoid any
coupling by this laser.

In any case  we have to shut off the laser by a mechanical
shutter in front of the trap for the observation of the revival
function in the ion cloud measurements. The applied power at 729
nm is measured by a beam pick-up photodiode at the entrance of
the trap. The instant when the power falls to zero is used to
fix the starting point for the fit of the revival function.

In the single ion experiments complete extinction of the laser at
729 nm is essential to avoid reduction of the length of the dark
intervals. We use a system of two synchronized mechanical
shutters, one at the entrance of the optical fiber and one at its
output, to guarantee maximum isolation from this radiation. When
both shutters are closed, the 729-nm light level is inferior to
the detection limit. For the excitation of quantum jumps, the
729-nm laser is applied to the laser-cooled ion, when the
fluorescence level falls below threshold, the shutters are
immediately closed. During the first measurement bins the 729-nm
radiation is thus still present in the trap and we have to remove
these earliest points to make sure that the decay conditions are
well defined. In our evaluation of the distribution of the
quantum jumps, we use data starting only at the fifth bin, which
means that we do not take into account the first 120 ms of a
quantum jump, and that quantum jumps whose duration is equal or
inferior to that value are completely omitted. As a consequence,
low-signal bins due to noise, which are typically of the length
of one bin are also eliminated.

\paragraph{866 nm}

\begin{figure}
\begin{flushleft}
   \epsfig{file=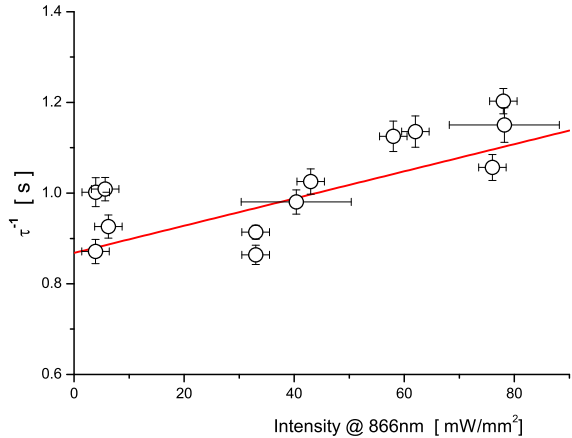, width=85mm}
  \end{flushleft}
  \caption{De-shelving effect due to the intensity of the 866 nm
  radiation. The y-error bar represents the 1$\sigma$-statistical uncertainty
  for the MLE of each point, while the x error bar stems from the
  uncertainty in the evaluation of the laser intensity seen by the
  ion.   For the extrapolation of the data to zero, x and y error bars are taken into account. }
  \label{f-deshelv866}
\end{figure}

The main effect which reduces the lifetime of the metastable state
is the coupling by the repumper laser at 866 nm \cite{block99},
which is necessary for laser-cooling.  We  insist on this point,
as this "helper" laser is often applied with optical power levels
largely beyond saturation to compensate for spatial and spectral
instabilities.

The determination of the laser intensities seen by the ions is
difficult to realize with high precision.
  On the one hand,  absolute calibration of a power meter is in general not constant in
time. More delicate still is the measure of the waist size in the
ultrahigh-vacuum vessel. We have measured the waist sizes on
different days by a commercial instrument (\textsl{Melles Griot
BeamAlyzer}) just in front of the entrance of the trap by
deviation with a high-quality mirror. This measurement is just an
estimate of the waist dimension, as the position and shape of the
focal point after the vacuum windows are certainly distorted.
Additionally, the trajectory of the ion may be smaller than the
waist of the laser. As a consequence, the error bar on the laser
intensity seen by a single ion is difficult to estimate and may
easily reach 10\%, and up to 30 \% for very low power levels (below 1$\mu W$).

We have checked our lifetime measurements versus the intensity of
the repumper laser (see figure \ref{f-deshelv866}),  observing a
reduction of the apparent lifetime of the $3D_{5/2}$-level due to
the AC Stark effect which couples the $3D_{5/2}$- and
$4P_{3/2}$-states. Data points are somewhat scattered, due to the
finite number of quantum jumps per run (cf section
\ref{s-dataan}) and the above discussed uncertainty in the power
measurement. The extrapolation to zero is made by a weighted
least-squares fit to all the measured data points taking into
account errors in both coordinates \cite{press1988}. On a
2$\sigma$-scale, all points but one fall into the confidence
level limits of the fit. As a matter of fact, the points at lower
power levels have a larger intensity error bar, and do thus contribute
less than may be expected from the linear representation.   The
measured lifetime of the ions varies as
 $  \tau_m^{-1} = \tau_{nat}^{-1}+ (3.0 \pm 0.6)\times
  10^{-3} I $
where \textit{I} is the intensity of the laser at 866 nm in
mW/mm$^2$, and $\tau_m$ and $\tau_{nat}$ are the measured and
natural lifetimes, respectively. The quantitative dependence of
the lifetime on the 866 nm-intensity we have found is of the same
order of magnitude as in previous measurements
\cite{block99,barton00}.

\subsection{Data analysis effects}
\label{s-dataan}

To determine the $3D_{5/2}$-level lifetime from the ion-cloud
experiments we fitted more than 1700 revival graphs using a
 least-squares fitting method. Because we could not
evidence any dependence of the lifetime on any experimental
parameter, the complete set of decay constants has been taken
into account to establish the final value.
It has been found for these values that different methods of data
analysis gave slightly different results. Actually, the mean
value of the fitted time constants is not identical to the fit of
the sum of the revival curves. Furthermore, harmonic and geometric
mean are not identical. We have simulated shelving data of the
ion cloud at a given decay rate having the same Gaussian noise
pattern as our experimental data. These generated data are free
from experimental bias such as ion loss or laser instabilities.
Fitting these data demonstrated the variations from the different
evaluation approaches, which could be as  large as 1.6 \% and
forms the major contribution to the total error bar.

For a single ion, the total number of quantum jumps taken into
account is superior to 40000. As has been discussed in section
\ref{s-single ion} the number of quantum jumps which can be
acquired in a one-day run is limited to roughly 10000. For
measurement times longer than this, temperature drifts start to
play a role in our actual experimental set-up, and ion loss may
also occur. For the experimental points in figure
\ref{f-deshelv866} the mean quantum jump number per set of
experimental parameters is of the order of 4000, giving a
statistical uncertainty of about 2-3\% per data set.  These
individual uncertainties together with the uncertainties on the power measurements are then
used to weight the data points in the extrapolation of the lifetime
values to zero repumper power, resulting in an overall statistical uncertainty of 1.8\%.

 Table \ref{table_deshelv} summarizes the quantitative
influence of the different effects which define the precision of
our measurements. Data analysis effects play a major role in the
evaluation of the electron-shelving curves in a small ion cloud,
whereas the main error contributions in single-ion measurements
come from the limited number of quantum jumps and the estimation
of the applied laser intensity independent of the data processing
method. The total error bars represent an uncertainty on the
measurement of the lifetimes of 2.5\% in the case of the ion
cloud and  1.5\% and 1.8\% for the single-ion experiments.

\section{Discussion and conclusions}

From the described measurements, the natural lifetime of the
$3d^2D_{5/2}$-level lifetime has been found to be $\tau_{IC} = (
1095 \pm 27)$ ms   in  an ion cloud and $\tau_{QJ} = (1152 \pm
20)$ ms in a single ion. These two values are very close, though
their 1$\sigma$ error bars do not overlap. This discrepancy could
be an evidence that ion-ion-collisions in the trap  eventually
contribute to the reduction of the lifetime. In the laser-cooled
ion cloud the particle density of the Ca$^+$-ions is roughly
$10^8 cm^{-3}$. If we suppose an eventual lifetime reduction on
the 1\%-level, we can estimate the upper limit of the
contribution of ion-ion collisions to be $\Gamma_{iic} \leq 1.2
\times 10^{-10}$ cm$^3$s$^{-1}$.

 As has been discussed in section \ref{s-single ion}, we
consider that the estimation of the lifetime in single ion
measurements is optimal when made by the method of maximum
likelihood. Nevertheless, we have also evaluated our data by
using the  least-squares minimization of the exponential fitting
curve and have found  (1088 $\pm$ 16) ms. This value is in very
good agreement with experimental values which have been
determined in the past  by least-squares methods
\cite{block99,lidberg99} with comparable number of events and
probability distributions. One of the causes generating  the
discrepancy between these earlier values and recent measurements
\cite{barton00,staanum03} as well as our value $\tau_{QJ} = (1152
\pm 20)$ ms could thus possibly  be  the choice of the data
processing method, which may account for differences of a couple
of percent depending on the number of considered quantum jumps
per run.

\begin{figure}
\begin{center}
   \epsfig{file=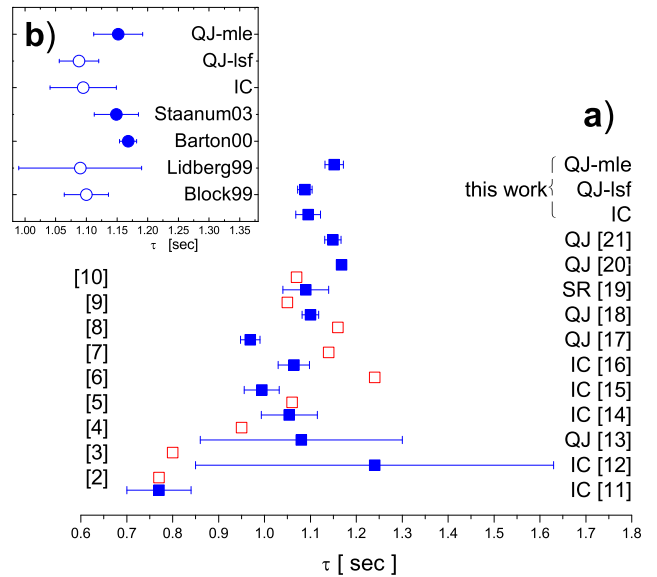, height=9cm}
  \end{center}
  \caption{a) Comparison of theoretical ($\square$) and experimental ($\blacksquare$) lifetimes for
   the $3D_{5/2}$-level of singly-ionized calcium, measurements
   have been made on ion clouds (IC), single ions (QJ), or in a
   storage ring (SR).
    b) Zoom of the six most recently measured lifetimes, represented with 2$\sigma$ error bars. The inset distinguishes the different
 evaluation approaches for the experimental data:  least-squares fitting (LSF) method to an exponential decay curve (\large$\circ$\small)
 and  estimate using the maximum-likelihood method (MLE)(\large$\bullet$\small).}
  \label{f-taucomp}
\end{figure}

Figure \ref{f-taucomp} shows all the measurements and calculations
that have been carried out on the Ca$^+$-ion over the last years.
A wide variety of theoretical models has been used to calculate
the lifetime of the metastable state. If we consider only the
more recent publications, the values still span over more than
20\% of the average lifetime value.
 Since 1999, experimental
measurements  are converging towards a value around 1100 ms.
Usually, lifetimes are given with a 1$\sigma$ error bar, which
corresponds to a poor confidence level. We have plotted the most
recent measurements with a $2\sigma$ error bar standing for a
95\% confidence level (see inset on figure \ref{f-taucomp}).
Values obtained in the storage ring \cite{lidberg99} suffer from
a strong collisional background and thus exhibit  the largest
error bars. The other values have been obtained on laser-cooled
single ions in linear and spherical traps. In the figure, a
graphical distinction has been made between values obtained by a
least-squares fit of the exponential decay curve or maximum
likelihood estimates.

 In
general, the overlap of the recent experimental results is very
good, and in particular, the agreement of our value with the one
from the {\AA}arhus group \cite{staanum03} is excellent.
The all-over precision of the experimental determination of the
metastable lifetime measurement is very high, making the Ca$^+$-ion
an ideal candidate for the comparison with
theoretical models.


In conclusion, we have measured the lifetime of the metastable
state to a 2.0\%-level and shown the influence of the data
processing method on the obtained value. Measurement of the $3d
^2D_{3/2}$-level lifetime would be interesting to allow direct
comparison of the fine-structure levels. The actual experimental
protocol does not allow such a measurement due to the coupling by
the repumper laser. A comparable metastable-level lifetime has
been measured in the Ba$^+$-ion \cite{yu97}. We plan to apply
this latter technique for the measurement of the $3D_{3/2}$
lifetime in Ca$^+$.

The high degree of control of our experimental parameters, that we
need for the preparation of the metrological project, has now
been achieved . The next step will be the localization of the ion
in the trap to better than a fraction of the emitted wavelength,
giving access to the Lamb-Dicke regime, where the first-order
Doppler broadening can be eliminated \cite{dicke53}.
Stabilization of the clock laser to an ULE cavity will be
required to make the laser linewidth compatible with the observed
ion features.

\begin{acknowledgements}
This work has benefitted from fruitful discussions with Peter
Staanum, David Lucas and Bruno Torr\'esani. 
 Our project has been financially
supported by the Bureau National de M\'etrologie.
\end{acknowledgements}

\end{document}